\documentclass[draft, onecolumn]{IEEEtran}
\usepackage{amsmath}
\usepackage{graphicx}
\usepackage{amsfonts}
\usepackage{amssymb}
\usepackage{mathrsfs}
\usepackage[mathscr]{euscript}
\usepackage{graphicx}
\usepackage{setspace}
\usepackage{tikz}

\makeatletter
\def\Ddots{\mathinner{\mkern1mu\raise\p@
		\vbox{\kern7\p@\hbox{.}}\mkern2mu
		\raise4\p@\hbox{.}\mkern2mu\raise7\p@\hbox{.}\mkern1mu}}
\makeatother

\newtheorem{definition}{Definition}
\newtheorem{theorem}{Theorem}
\newtheorem{lemma}{Lemma}

\newtheorem{example}{Example}
\newtheorem{remark}{Remark}

\begin{document}

\title{Complete $j$-MDP convolutional codes}

\author{P. J. Almeida$^{1}$ and Julia Lieb$^{2}$
\thanks{$^{1}$ Department of Mathematics, University of Aveiro, Portugal
	{\tt\small palmeida@ua.pt}}
\thanks{$^{2}$ Department of Mathematics, University of Aveiro, Portugal
	{\tt\small jlieb@ua.pt}}
}
\maketitle

\begin{abstract}
Maximum distance profile (MDP) convolutional codes have been proven to be very suitable for transmission
over an erasure channel. In addition, the subclass of complete MDP convolutional codes has the ability to restart decoding after a burst of erasures. However, there is a lack of constructions of these codes over fields of small size.
In this paper, we introduce the notion of complete $j$-MDP convolutional codes, which are a generalization of complete MDP convolutional codes, and describe their decoding properties. In particular, we present a decoding algorithm for decoding erasures within a given time delay $T$ and show that complete $T$-MDP convolutional codes are optimal for this algorithm. Moreover, using a computer search with the MAPLE software, we determine the minimal binary and non-binary field size for the existence of $(2,1,2)$ complete $j$-MDP convolutional codes and provide corresponding constructions. We give a description of all $(2,1,2)$ complete MDP convolutional codes over the smallest possible fields, namely $\mathbb F_{13}$ and $\mathbb F_{16}$ and we also give constructions for $(2,1,3)$ complete $4$-MDP convolutional codes over $\mathbb F_{128}$ obtained by a randomized computer search.
	
	\textbf{Keywords: Complete MDP convolutional codes, Erasure channel, complete $j$-MDP convolutional codes, minimum field size}
	
	\textbf{MSC: 94B10, 94B35}
\end{abstract}

\section{Introduction}	
Convolutional codes are especially suitable for transmitting over an erasure channel, which is the most used channel in multimedia traffic. An erasure channel is a communication channel where the receiver knows if a
received symbol is correct since symbols either arrive correctly or are erased. The Internet is an important example of an erasure channel.
The advantage of convolutional codes is the ability of considering a part of the sequence ("window") of any size and slide this window along the transmitted sequence depending on the erasures location to optimize the number of corrected erasures.

Crucial for the erasure correcting capability of a convolutional code are its column distances, which are limited by an upper bound, proven in \cite{RS99}. Convolutional codes attaining these bounds, i.e. convolutional codes whose column distances increase as rapidly as possible for as long as possible, are called maximum distance profile (MDP) convolutional codes. In \cite{vp}, the authors showed that MDP convolutional codes have optimal recovery rate. Moreover, they introduced the subclass of reverse MDP convolutional codes, which could recover more erasure patterns since they are also suitable for sliding backwards along the sequence. Finally, complete MDP convolutional codes, which are again a subclass of reverse MDP convolutional codes, have the additional benefit that they can correct even more erasure patterns than reverse MDP convolution codes, e.g. there is less waiting time when a large burst of erasures occurs and no correction is possible for some time (see \cite{vp}).

In this paper, we will introduce a new notion of convolutional codes, the so-called complete $j$-MDP convolutional codes, which contain the class of complete MDP convolutional codes. The column distances of these codes are maximal up to the $j$-th column distance. Moreover, they are also suitable for backward decoding and admit to restart the decoding after a burst of erasures if there is a window with sufficient small percentage of erasures after this burst. For small $j$, complete $j$-MDP convolutional codes have a weaker erasure correcting capability than complete MDP convolutional code but in turn are easier to construct. Moreover, if decoding has to be performed within time delay $T$, complete $T$-MDP convolutional codes are optimal and have the same performance as complete $j$-MDP codes with $j>T$ and hence also the same performance as complete MDP convolutional codes.

There are some general constructions for MDP (see \cite{strongly}, \cite{dr13}, \cite{dr16} and \cite{dr}) and complete MDP convolutional codes (see \cite{cmdp}). However, all of these constructions have the disadvantage that they only work over base fields of very large size.

This provokes the question for the minimal field size such that an MDP (respectively, complete MDP) convolutional code exists. Upper bounds on the necessary field size for the existence of MDP convolutional codes could be found in \cite{b} and \cite{nf}, for the existence of complete MDP convolutional codes in \cite{nf}. Furthermore, in \cite{nf}, lower bounds for the probability that a convolutional code is MDP (respectively complete MDP) are obtained.

For some special code parameters, the precise necessary field size as well as corresponding constructions are known. In \cite{bar}, $(n,n-1,2)$ MDP convolutional codes where $n\geq 4$ is a power of a $2$, are considered. In \cite{nf}, the author  considered $(n,1,1)$ (and $(n,n-1,1)$) MDP, reverse MDP and complete MDP convolutional codes and calculated the corresponding probabilities. In particular, \cite{nf} contains a description of all $(2,1,1)$ MDP, reverse MDP and complete MDP convolutional codes over the smallest possible field $\mathbb F_3$. In this paper, we continue this work giving a description of all $(2,1,2)$ complete MDP convolutional codes with minimal field size.

The paper is structured as follows. In Section 2, we start with some preliminaries about MDP, reverse MDP and complete MDP convolutional codes. In Section 3, we present a low delay erasure decoding algorithm for convolutional codes.
 In Section 4, we introduce complete $j$-MDP convolutional codes and their decoding capability. In Section 5, we study the $(2,1,2)$ complete $j$-MDP convolutional codes and, for each $j\leq 4$, we give the minimal binary and non-binary field sizes for the existence of a complete $j$-MDP convolutional code together with a corresponding construction. In particular, we give a complete description of all $(2,1,2)$ complete MDP convolutional codes over the smallest possible fields, namely $\mathbb F_{13}$ and $\mathbb F_{16}$. These results were obtained with the help of a computer search using the mathematical software Maple. In Section 6, we use a randomized computer search to find some examples of $(2,1,3)$ complete $4$-MDP convolutional codes and describe the computational difficulties for dealing with larger parameters. In Section 7, we describe the algorithms we used for the computer search. In Appendix, we include an illustrative sample of the computer programs we used.

\section{Preliminaries}

In this section, we summarize the basic definitions and properties concerning MDP convolutional codes.
One way to define a convolutional code is via polynomial generator matrices.

\begin{definition}
A \textbf{convolutional code} $\mathfrak{C}$ of \textbf{rate} $k/n$ is a free $\mathbb F[z]$-submodule of $\mathbb F[z]^n$ of rank $k$.
There exists $G(z)\in\mathbb F[z]^{n\times k}$ of full column rank such that
$$\mathcal{C}=\{v(z)\in\mathbb F[z]^n\ |\ v(z)=G(z)m(z)\ \text{for some}\ m(z)\in\mathbb F[z]^k\}.$$
$G(z)$ is called \textbf{generator matrix} of the code and is unique up to right multiplication with a unimodular matrix $U(z)\in Gl_k(\mathbb F[z])$.\\
The \textbf{degree} $\delta$ of $\mathcal{C}$ is defined as the maximal degree of the $k\times k$-minors of $G(z)$.
Let $\delta_1,\hdots, \delta_k$ be the column degrees of $G(z)$. Then, $\delta\leq\delta_1+\cdots+\delta_k$ and if $\delta=\delta_1+\cdots+\delta_k$, $G(z)$ is called a \textbf{minimal} generator matrix and $\max_{i\in\{1,\hdots,n\}}\delta_i$ is the \textbf{memory} of $\mathcal{C}$.\\
We refer to a convolutional code with rate $k/n$ and degree $\delta$ as $(n,k,\delta)$ convolutional code.
\end{definition}



There is a generic subclass of convolutional codes that could not only be described by an image representation via generator matrices but also by a kernel representation via the so-called parity-check matrices, which will be introduced in the following. Therefore, we need the notion of right prime and left prime polynomial matrices.

\begin{definition}
A polynomial matrix $G(z)\in\mathbb F[z]^{n\times k}$ with $k<n$ is called \textbf{right prime} if it has a left $k\times n$ polynomial inverse. For $k>n$, it is called \textbf{left prime} if it has a right $n\times k$ polynomial inverse.
\end{definition}


\begin{definition}
A convolutional code $\mathcal{C}$ is called \textbf{non-catastrophic} if one and therefore, each of its generator matrices is right prime.
\end{definition}



\begin{definition}
If $\mathfrak{C}$ is non-catastrophic, there exists a so-called \textbf{parity-check matrix} $H(z) \in\mathbb F[z]^{(n-k)\times n}$ of full rank, such that
$$\mathfrak{C} =\{ v(z)\in \mathbb F[z]^n\ |\ H(z)v(z) = 0 \in \mathbb F[z]^{n-k}\}.$$
\end{definition}

Throughout this paper we assume all convolutional codes to be non-catastrophic. It is always possible to choose a row reduced and left prime parity-check matrix and if $H(z)\in\mathbb F[z]^{(n-k)\times n}$ is a left prime and row reduced parity-check matrix of an $(n,k,\delta)$ convolutional code $\mathcal{C}$, then the sum of the row degrees of $H(z)$ is equal to $\delta$ (see \cite{ro01}).

%
%
%
%
%
%

We will need the representation by parity-check matrices to define complete MDP convolutional codes.
But first of all, we want to introduce MDP convolutional codes, for which we have to consider distances of convolutional codes. For this paper, the notion of column distances plays a crucial role because column distances are important for the decoding over an erasure channel. In this kind of channel, each symbol either arrives correctly or does not arrive at all.

\begin{definition}
The \textbf{Hamming weight} $wt(v)$ of $v\in\mathbb F^n$ is defined as the number of its nonzero components.\\
For $v(z)\in\mathbb F[z]^n$ with $\deg(v(z))=\gamma$, write $v(z)=v_0+\cdots+v_{\gamma}z^{\gamma}$ with $v_t\in\mathbb F^n$ for $t=0,\hdots,\gamma$ and set $v_t=0\in\mathbb F^n$ for $t\geq\gamma+1$. Then, for $j\in\mathbb N_0$, the \textbf{j-th column distance} of a convolutional code $\mathcal{C}$ is defined as
$$d_j^c(\mathcal{C}):=\min_{v(z)\in\mathcal{C}}\left\{\sum_{t=0}^j wt(v_t)\ |\ v_0\neq 0\right\}.$$
\end{definition}


There exist upper bounds for the column distances of a convolutional code.

\begin{theorem}\cite{RS99}\cite{strongly}\label{ub}
Let $\mathcal{C}$ be a convolutional code with rate $k/n$ and degree $\delta$. Then, it holds:\\
$d_j^c (\mathcal{C}) \leq (n-k)(j + 1) + 1$ for $j\in\mathbb N_0$
\end{theorem}

We are interested in convolutional codes whose column distances increase as rapidly as possible for as long as possible.

\begin{definition}\cite{mdp}
A convolutional code $\mathcal{C}$ of rate $k/n$ and degree $\delta$ has
  \textbf{maximum distance profile (MDP)} if
$$d_j^c(\mathcal{C})=(n-k)(j+1)+1\quad  \text{for}\ j=0,\hdots,L:=\left\lfloor\frac{\delta}{k}\right\rfloor+\left\lfloor\frac{\delta}{n-k}\right\rfloor$$
\end{definition}

According to \cite{strongly}, it is sufficient to have equality for $j=L$ in Theorem  \ref{ub} to get an MDP convolutional code.


In the following, we will provide criteria to check whether a convolutional code has a maximum distance profile. Therefore, we need the notion of trivially zero determinants.

\begin{definition}\label{sr}
Let $A=[\mu_{ij}]$ be a square matrix of order $m$ over $\mathbb{F}_{q^M}$ and $S_m$ the symmetric group of order $m$. Recall that the determinant of $A$ is given by

\begin{align}\label{deter}
|A|=\sum_{\sigma\in S_m}{sgn(\sigma)}\mu_{1\sigma(1)}\cdots \mu_{m\sigma(m)},
\end{align}

where the sign of the permutation $\sigma$, $sgn(\sigma)$, is $1$ (resp. $-1$) according to if $\sigma$ can be written as product of an even (resp. odd) number of transpositions. A {\em trivial term} of the determinant is a term of (\ref{deter}), $\mu_{1\sigma(1)}\cdots \mu_{m\sigma(m)}$, equal to zero. If $A$ is a square submatrix of a matrix $B$, with entries in $\mathbb{F}_{q^M}$, and all the terms of the determinant of $A$ are trivial we say that $|A|$ is a \textbf{trivial minor} of $B$.
\end{definition}

\begin{theorem}\label{cd}\cite{strongly}
Let $\mathcal{C}$ have a left prime parity-check $H(z)=\sum_{i=0}^{\nu}H_iz^i\in\mathbb F[z]^{n-k\times n}$. The following statements are equivalent:\\
(i) $d_j^c (\mathcal{C})=(n-k)(j + 1) + 1$ \\
(ii) $\mathcal{H}_j:=\left[\begin{array}{ccc} H_0 & & 0\\ \vdots & \ddots &  \\ H_j & \hdots & H_0 \end{array}\right]$ with $H_i\equiv 0$ for $i>\nu$ has the property that every full size minor that is not trivially zero, i.e. is formed by columns with indices $1\leq j_1<\cdots<j_{(j+1)(n-k)}$ with $j_{s(n-k)}\leq sn$ for $s=1,\hdots,j$, is nonzero.
\end{theorem}

From the preceding theorem, one can deduce the following decoding property.

\begin{theorem}\label{cd2}\cite{vp}
If $d_j^c (\mathcal{C})=(n-k)(j + 1) + 1$  and in any sliding window of length $(j + 1)n$ at most $(j+1)(n-k)$ erasures occur in a transmitted sequence, then complete recovery is possible by iteratively decoding the symbols ‘from left to right’.
\end{theorem}



%

We recall reverse MDP convolutional codes, which are advantageous for use in forward and backward decoding algorithms  \cite{vp}.

\begin{definition}\cite{h}
Let $\mathcal{C}$ be an $(n,k,\delta)$ convolutional code with right prime minimal generator matrix $G(z)$, which has entries $g_{ij}(z)$ and column degrees $\delta_1,\hdots,\delta_k$. Set $\overline{g_{ij}(z)}:=z^{\delta_j}g_{ij}(z^{-1})$. Then, the code $\overline{\mathcal{C}}$ with generator matrix $\overline{G(z)}$, which has $\overline{g_{ij}(z)}$ as entries, is also an $(n,k,\delta)$ convolutional code, which is called the \textbf{reverse code} to $\mathcal{C}$.\\
It holds: $v_0+\cdots+v_dz^d\in\overline{\mathcal{C}}\ \Leftrightarrow\ v_d+\cdots+v_0z^d\in\mathcal{C}$.
\end{definition}

\begin{definition}\cite{vp}
Let $\mathcal{C}$ be an MDP convolutional code. If $\overline{\mathcal{C}}$ is also MDP, $\mathcal{C}$ is called \textbf{reverse MDP} convolutional code.
\end{definition}

Reverse MDP convolutional codes can recover the same amount of erasures per window as MDP convolutional codes but not only from left to right but also from right to left, which allows in total to correct more erasure patterns, see \cite{vp}.

\begin{remark}\cite{vp}
Let $\mathcal{C}$ be an $(n,k,\delta)$ MDP convolutional code with $(n-k)\mid\delta$. Furthermore, let $H(z) = H_0 + \cdots +H_{\nu}z^{\nu}$ be a left prime and row proper parity-check matrix of $\mathcal{C}$. Then the reverse
code $\overline{\mathcal{C}}$ has parity-check matrix $\overline{H}(z) = H_{\nu} +\cdots +H_0z^{\nu}$. Therefore, $\mathcal{C}$ is reverse MDP if and only if every full size minor of the matrix
$$\widetilde{\mathcal{H}}_L:=\left[\begin{array}{ccc} H_{\nu} & \cdots & H_{\nu-L}\\  & \ddots & \vdots \\ 0 &  & H_{\nu} \end{array}\right]$$
formed from the columns with indices $j_1,\hdots,j_{(L+1)(n-k)}$
with $j_{s(n-k)+1} > sn$, for $s = 1,\hdots,L$ is nonzero.
\end{remark}

Next, we introduce complete MDP convolutional codes, which are even more advantageous for decoding than reverse MDP convolutional codes as there is less waiting time when a large burst of erasures occurs and no correction is possible for some time \cite{vp}.

\begin{definition}\cite{vp}\label{com}
Let $H(z)=H_0+H_1z+\cdots H_{\nu}z^{\nu}\in\mathbb F[z]^{(n-k)\times n}$ be a left prime parity-check matrix of the convolutional code $\mathcal{C}$ of rate $k/n$ and degree $\delta$. Set $L:=\left \lfloor\frac{\delta}{n-k}\right \rfloor+\left \lfloor\frac{\delta}{k}\right \rfloor$. Then
\begin{align}\label{ppc}
\mathfrak{H}:=\left(\begin{array}{ccccc}
H_{\nu} & \cdots & H_0 &   & 0 \\
  & \ddots &   & \ddots &   \\
0 &   & H_{\nu} & \cdots & H_0
\end{array}\right)  \in\mathbb F^{(L+1)(n-k)\times (\nu+L+1)n}
\end{align}
is called \textbf{partial parity-check matrix} of the code. Moreover, $\mathcal{C}$ is called \textbf{complete MDP} convolutional code if for any of its parity-check matrices $H(z)$, every full size minor of $\mathfrak{H}$ which is not trivially zero is nonzero.
\end{definition}

\begin{remark}
(i) Every complete MDP convolutional code is a reverse MDP convolutional code. \cite{vp}\\
(ii) A complete MDP convolutional code exists over a sufficiently large base field if and only if $(n-k)\mid\delta$. \cite{cmdp}
\end{remark}

As for $\mathcal{H}_L$ - when considering MDP convolutional codes - and additionally for $\widetilde{\mathcal{H}}_L$ - when considering reverse MDP convolutional codes - one could describe the not trivially zero full size minors of the partial parity-check matrix $\mathfrak{H}$ by conditions on the indices of the columns one uses to form the corresponding minor.

\begin{lemma}\cite{vp}\label{index}
A full size minor of $\mathfrak{H}$ formed by the columns $j_1,\hdots,j_{(L+1)(n-k)}$ is not trivially zero if and only if
\begin{itemize}
\item[(i)]
$j_{(n-k)s+1}>sn$
\item[(ii)]
$j_{(n-k)s}\leq sn+\nu n$
\end{itemize}
for $s=1,\hdots,L$.
\end{lemma}
%
%

In addition to the erasure correcting capability of reverse MDP convolutional codes, complete MDP convolutional codes admit the possibility to continue decoding after a window with too many erasures was received. This is described in the following theorem, which is an immediate consequence of the preceding lemma.

\begin{theorem}
If in a window of size $(\nu+L+1)n$
there are not more than $(L+1)(n-k)$ erasures, and if between position $1$ and $sn$ and
between position $(\nu+L+1)n$ and $(\nu+L+1)n-sn+1$
for $s = 1,\hdots,L + 1$, there are not more than $s(n-k)$
erasures, then full correction of all symbols in this interval is possible.
\end{theorem}

In case one receives one window that fulfills the properties of this theorem, one is able to recover all the erasures in this window, no matter what happens in the other windows. This might also help to recover the erasures that could not be recovered before.

The main advantage of convolutional codes is the use for sequential decoding with low delay. Here, backward decoding is not really helpful since it comes with a larger decoding delay. However, the possibility to restart the decoding after some symbols were lost is an important feature of complete MDP convolutional codes also for low delay decoding.

\section{A low delay decoding algorithm for convolutional codes}
In this section, we present a decoding algorithm for convolutional codes that recovers as many erasures as possible within a fixed decoding delay $T$.
Assume that for a codeword $v(z)=\sum_{i\in\mathbb N_0}v_iz^{i}$ of $\mathcal{C}$, the coefficients $v_0,\hdots,v_{t-1}$ arrive (correctly) for some $t\in\mathbb N_0$ and at least one component of the vector $v_t$ is erased. Let $H(z)=\sum_{i=0}^{\nu}H_iz^{i}$ be a parity-check matrix of $\mathcal{C}$. Then, for each $j\in\mathbb N_0$ and
\begin{align}\label{ppcj}
\mathfrak{H}_j:=\left(
  \begin{array}{ccccc} H_{\nu} & \cdots & H_0 & &  0\\
     & \ddots & & \ddots & \\
    0 & & H_{\nu} & \cdots & H_0 \\
    \end{array}
\right)\in \mathbb F^{(j+1)(n-k)\times (\nu+j+1)n },
\end{align}
one has $\mathfrak{H}_j [v_{t-\nu}, \hdots ,v_{t+j}]=\mathbf{0}$, where $v_i=0$ for $i\notin\{0,\hdots,\deg(v)\}$. 
Consequently, to recover the erasures in $v_{t},\hdots, v_{t+j}$, one has to solve the system of linear euqations
\begin{align}\label{system}
\mathfrak{H}_j^{(e)}[v^{(e)}_t, \hdots ,v^{(e)}_{t+j}]=-\mathfrak{H}_j^{(r)}[v^{(r)}_t, \hdots ,v^{(r)}_{t+j}]
\end{align}
where $v^{(e)}_i$ and $v^{(r)}_i$ denote the erased and received components of $v_i$, respectively, and $\mathfrak{H}_j^{(e)}$ and $\mathfrak{H}_j^{(r)}$ denote the corresponding columns of $\mathfrak{H}_j$. The erasures $[v^{(e)}_t, \hdots ,v^{(e)}_{t+j}]$ are recovered if and only if the system has a unique solution, i.e. if and only if $\mathfrak{H}_j^{(e)}$ has full column rank.

Let $T$ be the maximal delay one can or wants to tolerate during the recovering process of the erasures of a received sequence. We will present an algorithm for general convolutional codes that recovers all erasures within a delay of $T$ if the received erasure pattern is such that this is possible. Moreover, the algorithm recovers the first not yet corrected erasures with minimal possible delay.

There are two main variations of the decoding process. Either one assumes that one knows which erasure patterns can be corrected with the code one uses (this has then to be determined before but it is the same for every use of the same code), or one has to check during the decoding which erasure patterns can be corrected.
Knowing what can be corrected means that for the used code, we know for each $j$, which submatrices of $\mathfrak{H}_j$ have full column rank.\\

\textbf{Algorithm}\\
\textbf{1}: Set $i=-1$ and $c=-(\nu+1)$.\\
\textbf{2}: Repeat $i=i+1$ until $\hat{v}_i$ contains erasures or the end of the transmitted sequence is reached.\\
\textbf{3}: Set $j=0$.\\
\textbf{4}: Set $t=\max(i-\nu, c)$ and $s=i+j-t-\nu$. If $v_i$ can be recovered solving the linear system of equations induced by $\mathfrak{H}_s$ and $v_t,\hdots,v_{i-1}, v_i,\hdots,v_{i+j}$, go to 6, otherwise go to 5.\\
\textbf{5}: If $j=T$, we cannot recover the erasures in $v_i$, set $c=i$ and go back to 2. Otherwise, we set $j=j+1$ and go back to 4.\\
\textbf{6}: Recover the erasures in $v_i$ (and if possible also erasures in $v_{i+1},\hdots, v_{i+j}$), solving the system of linear equations \eqref{system}. Replace the erased symbols with the correct symbols and go back to 2.\\

The algorithm can be explained as follows: We first try to recover the first window of size $n$ that contains erasures as fast as possible. If this is not possible within delay $T$, these erased components have to be declared as lost. We move to the next window but since we have not recoverable erasures before, we need to consider the matrix $\mathfrak{H}_s$ for forming the linear system of equations that is used for decoding.

\medskip

It is also possible to correct part of the erasures of one coefficient vector of the polynomial codeword vector, even if not all of its erasures can be recovered. It depends on the corresponding application if it makes sense to do this or not.
To know before if partly correction is possible, one would need to know if there are columns of $\mathfrak{H}_j^{(e)}$ that are not in the span of the other columns of $\mathfrak{H}_j^{(e)}$.

\medskip

The aim of the following section is to construct convolutional codes with an optimal performance for the presented algorithm. It turns out that the optimal choice are so-called complete j-MDP codes for $j=T$, which will be introduced in the following section.

\section{Complete $j$-MDP convolutional codes}


It turns out to be hard to find constructions of MDP and especially complete MDP convolutional codes over fields of small size. In this section, we introduce a new class of codes, the so-called complete $j$-MDP convolutional codes, which are easier to obtain than complete MDP convolutional codes and are optimal for bounded delay decoding.

\begin{definition}
Let $(n-k)\mid\delta$ and $\mathcal{C}$ be an $(n,k,\delta)$ convolutional code with left prime parity-check matrix $H(z) = H_0 + \cdots +H_{\nu}z^{\nu}$ where $\nu=\frac{\delta}{n-k}$.
For $j=0,\hdots,L$, define $\mathfrak{H}_j$ as in \eqref{ppcj}
We call $\mathcal{C}$ a \textbf{complete $j$-MDP convolutional code} if all fullsize minors of $\mathfrak{H}_j$ that are not trivially zero are nonzero.
\end{definition}

Note that according to Definition \ref{com}, a complete $L$-MDP convolutional code is the same as a complete MDP convolutional code. Moreover, it follows from Theorem \ref{cd} that the colummn distances of a complete $j$-MDP convolutional code reach the upper bound of Theorem \ref{ub} up to the $j$-th column distance.

The following lemma is an easy consequence of Lemma \ref{index}.

\begin{lemma}\label{in2}
A full size minor of $\mathfrak{H}_j$ formed by the columns $j_1,\hdots,j_{(j+1)(n-k)}$ is not trivially zero if and only if
\begin{itemize}
\item[(i)]
$j_{(n-k)s+1}>sn$
\item[(ii)]
$j_{(n-k)s}\leq sn+\nu n$
\end{itemize}
for $s=1,\hdots,j$.
\end{lemma}

Furthermore, the following lemma implies that the number of erasure patterns that could be corrected with an $(n,k,\delta)$ complete $j$-MDP convolutional code and the necessary field size for the existence of such a code increase with $j$.

\begin{lemma}\label{less}
 If for some $i=0,\hdots,L$, $\mathcal{C}$ is a complete $i$-MDP convolutional code, then $\mathcal{C}$ is also a complete $j$-MDP convolutional code for all $j\leq i$.
\end{lemma}

\underline{Proof:}\\
Without loss of generality, we could assume $i=j+1$ (the rest follows per induction).
Let $a\in\mathbb F$ be any not trivially zero fullsize minor of $\mathfrak{H}_j$ and $A\in\mathbb F^{(j+1)(n-k)\times(j+1)(n-k)}$ be the matrix consisting of the corresponding columns of $\mathfrak{H}_j$, i.e. $\det(A)=a$. Denote the indices of the columns of $\mathfrak{H}_j$ that were used to form $A$ by $l_1,\hdots,l_{(j+1)(n-k)}$.
Consider the fullsize minor of $\mathfrak{H}_i$ formed by the union of columns with indices $l_1,\hdots,l_{(j+1)(n-k)}$ and of $n-k$ of the last $n$ columns of $\mathfrak{H}_i$. This fullsize minor of $\mathfrak{H}_i$, which we denote by $b$, is not trivially zero and as $\mathcal{C}$ is a complete $i$-MDP convolutional code, it is nonzero. Moreover, it holds $b=a\cdot\det(D)$ where $D$ is a $(n-k)\times(n-k)$ submatrix of $H_0$. Hence $a\neq 0$ and consequently, $\mathcal{C}$ is a complete $j$-MDP convolutional code.
\hfill$\blacksquare$\\

The definition of complete $j$-MDP convolutional codes as well as the preceding lemma imply that the class of complete $j$-MDP convolutional codes contains the complete MDP convolutional codes, but it neither contains the reverse MDP convolutional codes nor is it contained in the class of MDP convolutional codes. However, complete $j$-MDP convolutional codes are optimal for bounded delay decoding as the following theorem states.

\begin{theorem}\label{best}
Complete $j$-MDP convolutional codes are optimal for sequential erasure decoding with maximal delay $T=j$ (using the algorithm of the  preceding section), i.e. if an erasure pattern cannot be corrected within delay $j$ by a complete $j$-MDP convolutional code, it cannot be corrected by any other convolutional code with the same parameters.
\end{theorem}

\underline{Proof:}\\
Decoding within delay $j$ means solving linear systems of equations whose coefficient matrices are submatrices of $\mathfrak{H}_s$ with $s\leq j$; see step 4 of the algorithm. By definition and the preceding lemma, complete $j$-MDP convolutional codes have the property that all these coefficient matrices for which this is possible have full column rank. Hence, they are optimal for this kind of decoding.
\hfill$\blacksquare$\\

The following theorem describes which erasure patterns can be corrected by a complete $j$-MDP convolutional code.

\begin{theorem}\label{dec}
If $\mathcal{C}$ is a complete $j$-MDP convolutional code, then it has the following decoding properties:\\
\textbf{(1)} If in any sliding window of length $(j + 1)n$ at most $(j + 1)(n-k)$ erasures occur, complete recovery is possible.\\
\textbf{(2)} If in a window of size $(\nu + j + 1)n$ there are not more than
$(j + 1)(n-k)$ erasures, and if between position 1 and sn and
between position $(\nu + j + 1)n$ and $(\nu + j + 1)n-sn+1$, for
$s = 1,...,j$, there are not more than $s(n-k)$ erasures,
then full correction of all symbols in this interval is possible.
\end{theorem}

\underline{Proof:}\\
Property (1) is a consequence of Theorem \ref{cd} and Theorem \ref{cd2} and property (2) follows from Lemma \ref{in2}.
\hfill$\blacksquare$\\

In the following section, we will apply the results of this section to the $(2,1,2)$ complete $j$-MDP convolutional codes that we will construct there.

\section{Construction of $(2,1,2)$ complete $j$-MDP convolutional codes}

The aim of this section is the construction of $(2,1,2)$ complete $j$-MDP convolutional codes over fields with minimum possible field size. For these code parameters, we have $\nu=2$, $L=4$ and the following partial parity-check matrix:

\begin{equation}\label{Matrix5_14}
\mathfrak{H}:=\left(\begin{array}{ccccccc}
H_{2} & H_1 & H_0 & & & &  \\
& H_2 & H_1  & H_0 & & & \\
& & H_2 & H_1  & H_0 & &\\
& & & H_2 & H_1  & H_0 & \\
& & & & H_2 & H_1  & H_0
\end{array}\right)  \in\mathbb F^{5\times 14}.
\end{equation}

where $H_i\in\mathbb F^{1\times 2}$ for $i=0,1,2$. To construct complete $j$-MDP convolutional codes, one needs left prime parity-check matrices. As $H(z)=[h_1(z)\ h_2(z)]$ with $h_1(z), h_2(z)\in\mathbb F[z]$, $H(z)$ is left prime if and only if $h_1(z)$ and $h_2(z)$ are coprime. It is well known that this is true if and only if the resultant 
$$Res(h_1,h_2):=\det\left[\begin{array}{cccc}
h_{1,2} & h_{1,1} & h_{1,0} & 0\\
0 & h_{1,2} & h_{1,1} & h_{1,0}\\
h_{2,2} & h_{2,1} & h_{2,0} & 0\\
0 & h_{2,2} & h_{2,1} & h_{2,0}
\end{array}\right]=\det\left[\begin{array}{cccc}
h_{1,0} & h_{2,0} & 0 & 0\\
 h_{1,1} & h_{2,1} & h_{1,0} & h_{2,0}\\
h_{1,2} & h_{2,2} & h_{1,1} & h_{2,1}\\
0 & 0 & h_{1,2} & h_{2,2}
\end{array}\right]$$
 is nonzero.
For $j\geq 3$, $Res(h_1,h_2)$ is a factor of a nontrivial fullsize minor of $\mathfrak{H}_j$. Hence, that all nontrivial fullsize minors of $\mathfrak{H}_j$ are nonzero implies that the resultant is nonzero. Therefore, it is enough to consider the nontrivial fullsize minors of $\mathfrak{H}_j$ to show complete $j$-MDP without having to take care of the resultant. 

\begin{remark}
In \cite{gj}, it is shown that for $(n,k,\delta)$ convolutional codes with $(n-k)\mid\delta$, if the nontrivial fullsize minors of $\mathcal{H}_L$ are nonzero, then the parity-check matrix is left prime. The proof presented there can be easily adapted to show that if also $k\mid\delta$, then it is enough to consider $\mathcal{H}_{L-1}$, i.e. if all nontrivial fullsize minors of $\mathcal{H}_{L-1}$ are nonzero, than $H$ is left prime. As $\mathcal{H}_{L-1}$ is a submatrix of $\mathfrak{H}_{L-1}$, this shows in a more general setup that for $(2,1,2)$ complete $j$-MDP convolutional codes with $j\geq 3$ we do not need to consider the resultant (which of course can be seen easier for this special parameters as mentioned before).
\end{remark}

However, for $j\leq 2$, we have to consider the resultant in addition to the fullsize minors of $\mathfrak{H}_j$ for the construction of $(2,1,2)$ complete $j$-MDP convolutional codes.

\begin{remark}\label{n}
Notice that two left prime matrices $H(z), \hat{H}(z)\in\mathbb F[z]^{2\times 1}$ are parity-check matrices of the same $(2,1,2)$ convolutional code if and only if there exists $a\in \mathbb{F}\setminus\{0\}$ such that $H(z)=a\hat{H}(z)$. 
\end{remark}

In the following section, for all $j\in\{0,\hdots 4\}$, we will determine the smallest field as well as the smallest binary field over which a $(2,1,2)$ complete $j$-MDP convolutional code exists and give constructions for such codes.


\subsection{$(2,1,2)$ complete $0$-MDP convolutional codes}

It holds $\mathfrak{H}_0=[H_2\ H_1\ H_0]$ and obviously $\mathfrak{C}$ is complete $0$-MDP if and only if all entries of the vectors $H_i$ for $i\in\{0,1,2\}$ are nonzero and $Res(h_1,h_2)\neq 0$. Such a code does not exist over $\mathbb F_2$ since the parity-check matrix with $H_i=[1\ 1]$ for $i\in\{0,1,2\}$ is not left prime. Actually $H(z)=[z^2+z+1\ z^2+z+1]$ and $\hat{H}=[1\ 1]$ are parity-check matrices of the same code, which is a $(2,1,0)$ convolutional code, i.e. in fact a $[2,1]$ block code. This illustrates that the degree of a convolutional code can only be obtained from a parity-check matrix that is left prime. For all finite fields with $|\mathbb F|>2$, the code with parity-check matrix $H(z)$ where $H_2=[1\ 1]$, $H_1=[1\ \alpha]$ $H_0=[1\ 1]$ and $\alpha\in\mathbb F\setminus\{0,1\}$ is a $(2,1,2)$ complete $0$-MDP convolutional code.

According to Theorem \ref{dec} such a code has the following decoding properties: If in every window of size $2$ there is not more than $1$ erasure, full recovery is possible. Moreover, if this condition is violated at some point but later there is window of size $6$ with at most $1$ erasure, this erasure can be recovered and there is enough guard space to start again with the decoding. This means that full recovery is possible if every second symbol is erased but any burst of two or more erasures cannot be corrected. This should be illustrated with the following example.

\begin{example}
Assume that we have the following erasure pattern where $x$ denotes an erasure and $\surd$ a correctly received symbol\\
\begin{tabular}{| c | c  | c  |c |c | c |c |c|c|}
\hline
  x \ $\surd$ & $\surd$ \ x & x\ $\surd$ & x \ x & x \ $\surd$ & $\surd$ \ $\surd$ & $\surd$ \ $\surd$ & x \ $\surd$ & $\surd$ \ x \\
    \hline
\end{tabular}\\

In the first three windows of size two the erasures could be recovered but not the two erasures in the fourth window. However, after these two erasures, which have to be declared as lost, there is a window of size 6 with only $1$ erasure, which could be recovered. Moreover, this enables us to start the decoding again and to recover also the $2$ erasures in the last two windows of size $2$.
\end{example}

In the following sections, we will consider $(2,1,2)$ complete $j$-MDP convolutional codes with $j\geq 1$. According to Lemma \ref{less} these codes are also complete $0$-MDP and hence we know that all entries of the coefficients of $H(z)$ have to be nonzero.

\subsection{$(2,1,2)$ complete $1$-MDP and complete $2$-MDP convolutional codes}

In this section, we treat both $(2,1,2)$ complete $1$-MDP and $(2,1,2)$ complete $2$-MDP convolutional codes since the minimum binary field size for their existence is the same, which will be shown in the following.

\begin{theorem}
If $|\mathbb F|\leq 4$, there exists no $(2,1,2)$ complete $1$-MDP and hence also no $(2,1,2)$ complete $2$-MDP convolutional code over $\mathbb F$.
\end{theorem}

\underline{Proof:}\\
For the existence of a $(2,1,2)$ complete $1$-MDP convolutional code over $\mathbb F$ it is necessary to have four pairwise linearly independent vectors in $\mathbb F^2$ where additionally all entries of these vectors have to be nonzero. This is not possible if $|\mathbb F|\leq 4$.
\hfill$\blacksquare$\\

Hence the minimum binary field size for a $(2,1,2)$ complete $1$-MDP convolutional code and consequently also for a $(2,1,2)$ complete $2$-MDP convolutional code is at least $8$.

Some of the results in this and the next section were obtained with the help of computer search using Maple, so we will not write a formal proof. An explanation of the algorithms used is given in section \ref{Maple}. The following is one of these results:

\begin{theorem}
There exist $714\times 7^2$
values for $[H_2\ H_1\ H_0]\in(\mathbb F_{8}\setminus\{0\})^6$ such that $H(z)=\sum_{i=0}^2H_iz^{i}$ is the left prime parity-check matrix of an $(2,1,2)$ complete $1$-MDP convolutional code over $\mathbb F_{8}$, i.e. with Remark \ref{n} one has $714\times 7$ such codes. Moreover, there exist $126\times 7^2$ values for $[H_2\ H_1\ H_0]\in(\mathbb F_{8}\setminus\{0\})^6$ such that $H(z)=\sum_{i=0}^2H_iz^{i}$ is the parity-check matrix of an $(2,1,2)$ complete $2$-MDP convolutional code over $\mathbb F_{8}$, i.e. with Remark \ref{n} one has $126\times 7$ such codes. An example for such a code is given by $[H_2\ H_1\ H_0]=[1\ 1\ 1\ \alpha\ \alpha+1\ \alpha+1]$ where $\alpha$ is a primitive element of $\mathbb F_8$.
\end{theorem}


To determine the minimum (possibly non binary) field size for the existence of such codes, one has to check the existence over $\mathbb F_5$ and $\mathbb F_7$. Again with the help of computer search using Maple, we obtain the following two theorems.

\begin{theorem}
There exist $20\times 4^2$
values for $[H_2\ H_1\ H_0]\in(\mathbb F_{5}\setminus\{0\})^6$ such that $H(z)=\sum_{i=0}^2H_iz^{i}$ is the left prime parity-check matrix of an $(2,1,2)$ complete $1$-MDP convolutional code over $\mathbb F_{5}$, i.e. with Remark \ref{n} one has $20\times 4$ such codes. An example for such a code is given by $[H_2\ H_1\ H_0]=[1\ 1\ 1\ 2\ 1\ 2]$.
\end{theorem}

According to Theorem \ref{dec} an $(2,1,2)$ complete $1$-MDP convolutional code has the following decoding properties:\\
(1) If in every window of size 4, there are at most 2 erasures, then complete recovery is possible (forward and backward).\\
(2) If in one window of size 8, there are at most 2 erasures (and they are distributed according to the conditions of Theorem \ref{dec}, i.e. not too many at the edges), then full recovery in this window is possible.

\begin{theorem}
There exists no $(2,1,2)$ complete $2$-MDP convolutional code over $\mathbb F_{5}$. Moreover, there exist $14\times 6^2$
values for $[H_2\ H_1\ H_0]\in(\mathbb F_{7}\setminus\{0\})^6$ such that $H(z)=\sum_{i=0}^2H_iz^{i}$ is the left prime parity-check matrix of an $(2,1,2)$ complete $2$-MDP convolutional code over $\mathbb F_{7}$, i.e. with Remark \ref{n} one has $14\times 6$ such codes. An example for such a code is given by $[H_2\ H_1\ H_0]=[1\ 1\ 1\ 2\ 5\ 5]$.
\end{theorem}

According to Theorem \ref{dec} an $(2,1,2)$ complete $2$-MDP convolutional code has the following decoding properties:\\
(1) If in every window of size 6, there are at most 3 erasures, then complete recovery is possible (forward and backward).\\
(2) If in one window of size 10, there are at most 3 erasures (and they are distributed according to the conditions of Theorem \ref{dec}, i.e. not too many at the edges), then full recovery in this window is possible.\\

We want to compare the performances of $(2,1,2)$ complete $1$-MDP and $(2,1,2)$ complete $2$-MDP convolutional codes with the performances of $(2,1,2)$ MDP and $(2,1,2)$ reverse MDP convolutional codes. We already saw that within delay $j$ complete $j$-MDP convolutional codes are optimal but also for larger delay, they can be compared to MDP and reverse MDP convolutional codes. The smallest field over which a $(2,1,2)$ MDP convolutional code could exist is $\mathbb F_7$
(see \cite{b}). Such a code could correct forward if in each window of size 10 there are at most 5 erasures. As each reverse MDP convolutional code is an MDP convolutional code, the minimum field size for a $(2,1,2)$ reverse MDP convolutional code is at least $7$. Such a code could correct forward and backward if in each window of size 10 there are at most 5 erasures, see \cite{vp}.\\
It is possible to find erasures patterns that a $(2,1,2)$ complete $1$-MDP and hence also a $(2,1,2)$ complete $2$-MDP convolutional code can correct and a $(2,1,2)$ reverse MDP and hence also a $(2,1,2)$  MDP convolutional code not.

\begin{example}
The following erasure pattern could be completely corrected with a $(2,1,2)$ complete $1$-MDP convolutional code but not with a $(2,1,2)$ reverse MDP convolutional code:\\
\begin{tabular}{|  c| c | c  | c  |c |c | c |c |c|c|c|c|c|}
\hline
 x  \ x & x \ $\surd$ & $\surd$ \ x & x\ $\surd$ & x \ $\surd$ & x \ $\surd$ & $\surd$ \ $\surd$ & $\surd$ \ $\surd$ & x \ $\surd$ & $\surd$ \ x & x\ $\surd$ & x \ $\surd$ & x\ x\\
 \hline
 1 & 2 & 3 & 4 & 5 & 6 & 7 & 8 & 9 & 10 & 11 & 12 & 13\\
    \hline
\end{tabular}\\
\normalsize

Again $x$ denotes an erasure and $\surd$ a correctly received symbol and the second line of the table numbers the windows of size $2$.
\end{example}

\underline{Proof:}\\
As the union of the first five windows of size 2 forms a window of size $10$ with $6$, i.e. more than $5$, erasures this pattern cannot be corrected forward with a $(2,1,2)$ MDP convolutional code. As also the union of the last five windows of size $2$ contains more than $5$ erasures, backward decoding with a $(2,1,2)$ reverse MDP convolutional code is also not possible.\\
However, correction with a $(2,1,2)$ complete $1$-MDP convolutional code is possible with the following steps:
\begin{enumerate}
\item The union of the size-$2$-windows $5$ to $8$ forms a window of size $8$ with $2$ erasures, which can be recovered.
\item As the erasures in windows $5$ and $6$ are already recovered it is possible to recover the erasures in windows $3$ and $4$ with backward decoding (window of size $4$ with $2$ erasures).
\item After steps $1$ and $2$, the union of the size-$2$-windows $2$ to $5$ forms a window of size $8$ with $1$ erasure, which can be recovered.
\item Windows $1$ and $2$ form now a window of size $4$ with $2$ erasures that can be recovered with forward decoding.
\item Windows $9$ and $10$ form a window of size $4$ with $2$ erasures that can now be recovered with forward decoding.
\item Windows $11$ and $12$ form a window of size $4$ with $2$ erasures that can now be recovered with forward decoding.
\item Windows $12$ and $13$ form now a window of size $4$ with $2$ erasures that can now be recovered with backward decoding.
\end{enumerate}
For further explanation about sliding window decoding with MDP, reverse MDP and complete MDP convolutional codes, see \cite{vp}.
\hfill$\blacksquare$\\

On the other hand, there are also erasure patterns that can be corrected by a $(2,1,2)$ MDP convolutional code but not by a $(2,1,2)$ complete $2$-MDP convolutional code but they are harder to find and moreover, according to Theorem \ref{best}, they can only exist if one can tolerate a delay larger than $j=2$.

\begin{example}
The following erasure pattern could be corrected with a $(2,1,2)$ MDP convolutional code if the tolerated decoding delay is at least $4$ but not with a $(2,1,2)$ complete $2$-MDP convolutional code:\\
\begin{tabular}{|  c| c | c  | c  |c |}
\hline
 x  \ x & x \ $\surd$ & x\ x & $\surd$ \ $\surd$ & $\surd$ \ $\surd$ \\
    \hline
\end{tabular}\\
\normalsize
\end{example}

\underline{Proof:}\\
As we have a window of size $10$ with $5$ erasures, erasure recovery is possible with a $(2,1,2)$ MDP convolutional code with delay $4$.\\
However, neither decoding property (1) nor decoding property $(2)$ of a $(2,1,2)$ complete $2$-MDP convolutional code enable any recovery.
\hfill$\blacksquare$\\

\subsection{$(2,1,2)$ complete $3$-MDP and complete MDP convolutional codes}

The results in this section were all obtained with the help of Maple. We found the following:

\begin{theorem}
For $r\leq 3$, there exists no $(2,1,2)$ complete $3$-MDP convolutional code over $\mathbb F_{2^r}$.
\end{theorem}

The preceding theorem implies that the minimum binary field size for an $(2,1,2)$ complete MDP convolutional code is at least $16$. The following theorem states that it is exactly $16$ and gives a complete description of all such codes.

\begin{theorem}
There are $600\times 15$ $(2,1,2)$ complete $3$-MDP convolutional codes over $\mathbb F_{2^4}$ and a $(2,1,2)$ convolutional code over $\mathbb F_{2^4}$ is complete MDP if and only if
$$H_2=[\beta\ \gamma],\ H_1=[\beta\alpha^{i_1}\ \gamma\alpha^{i_2}],\ H_0=[\beta\alpha^{i_3}\ \gamma\alpha^{i_4}]$$
where $\alpha$ is a primitive element of $\mathbb F_{2^4}$,
$\beta, \gamma\in\mathbb F_{2^4}\setminus\{0\}$,
$i_1\in\{0,\hdots,14\}$,\\
$i_2= i_1+3k \mod 15$ for $k=1,2,4,8$,\\
$i_3=i_4=i_1+i_2-4^jk \mod 15$ for $j=0,1$.\\
Therefore, there are exactly $15^2\times 120$ values for $[H_2\ H_1\ H_0]\in(\mathbb F_{2^4}\setminus\{0\})^6$ and so there are exactly $15\times 120$ $(2,1,2)$ complete MDP convolutional codes over $\mathbb F_{2^4}$.
\end{theorem}

Two possibilities are given in the following example.

\begin{example}
If $\beta=\gamma=1, i_1=0, k=1$ and $j=0$ we obtain $H_2=[1\ 1]$, $H_1=[1\ \alpha^{3}]$, $H_0=[\alpha^{2}\ \alpha^{2}]$.

If $\beta=\gamma=1, i_1=2, k=8$ and $j=1$ we obtain $H_2=[1\ 1]$, $H_1=[\alpha^2\ \alpha^3+\alpha^{2}+\alpha]$ and $H_0=[\alpha^3+\alpha^{2}+\alpha\ \alpha^3+\alpha^{2}+\alpha]$.
\end{example}

Next, we present the minimum size of a (finite) prime field $\mathbb F_{p}$ for which we have a $(2,1,2)$ complete MDP convolutional code. We were again able to find a complete description of all such codes.

\begin{theorem}
There exists no $(2,1,2)$ complete $3$-MDP or complete MDP convolutional code over $\mathbb F_{q}$, for $q=p^r\leq 11$ an odd prime power.  Moreover, there are $240\times 12$ $(2,1,2)$ complete $3$-MDP convolutional codes over $\mathbb F_{13}$ and a $(2,1,2)$ convolutional code over $\mathbb F_{13}$ is complete MDP if and only if
$$H_2=[\beta\ \gamma],\ H_1=[\beta 2^{i_1}\ \gamma 2a^{i_2}],\ H_0=[\beta 2^{i_3}\ \gamma 2^{i_4}]$$
where $2$ is a primitive element of $\mathbb F_{13}$,
$\beta, \gamma\in\mathbb F_{13}\setminus\{0\}$,
$i_1\in\{0,\hdots,11\}$,\\
$i_2= i_1+6 \mod 12$,\\
$i_3=i_4=2i_1+1+6j \mod 12$ for $j=0,1$.\\
Therefore there are exactly $12^2\times 24$ values for $[H_2\ H_1\ H_0]\in(\mathbb F_{13}\setminus\{0\})^6$ and so there are exactly $12\times 24$ $(2,1,2)$ complete MDP convolutional codes over $\mathbb F_{13}$.
\end{theorem}

\section{$(2,1,3)$ complete $j$-MDP convolutional codes}\label{comp213}

In this section, we explain the computational difficulties while dealing with, for example, a $(2,1,3)$ complete $j$-MDP convolutional code and exhibit a few examples of this type of codes.
  
The computational effort is already enormous for a $(2,1,3)$ complete $j$-MDP convolutional code, since we need to find $H_2=[a,b]$, $H_1=[c,d]$, $H_0=[e,f]$ (here we can take $H_3=[1,1]$). There are $15^6=11 390 625$ different possibilities if we consider $\mathbb{F}_{16}$ and $31^6=887 503 681$ if we consider $\mathbb{F}_{32}$ (but we believe the smallest binary finite field is at least $\mathbb{F}_{64}$, although we were unable to find an example by a randomized computer search).  

For these code parameters, we have $\nu=3$, $L=6$ and the following partial parity-check matrix:

\begin{equation}\label{Matrix7_20}
\mathfrak{H}:=\left(\begin{array}{cccccccccc}
H_3 & H_{2} & H_1 & H_0 & & & & & &  \\
& H_3 & H_2 & H_1  & H_0 & & & & & \\
& & H_3 & H_2 & H_1  & H_0 & & & & \\
& & & H_3 & H_2 & H_1  & H_0 & & & \\
& & & & H_3 & H_2 & H_1  & H_0 & & \\
& & & & & H_3 & H_2 & H_1  & H_0 & \\
& & & & & & H_3 & H_2 & H_1  & H_0 
\end{array}\right)  \in\mathbb F^{7\times 20}.
\end{equation}

where $H_i\in\mathbb F^{1\times 2}$ for $i=0,1,2,3$. Again, to construct complete $j$-MDP convolutional codes, one needs left prime parity-check matrices. As $H(z)=[h_1(z)\ h_2(z)]$ with $h_1(z), h_2(z)\in\mathbb F[z]$, $H(z)$ is left prime if and only if $h_1(z)$ and $h_2(z)$ are coprime. It is well known that this is true if and only if the resultant 
$$Res(h_1,h_2):=\det\left[\begin{array}{cccccc}
h_{1,3} & h_{1,2} & h_{1,1} & h_{1,0} & 0 & 0\\
0 & h_{1,3} & h_{1,2} & h_{1,1} & h_{1,0} & 0\\
0 & 0 & h_{1,3} & h_{1,2} & h_{1,1} & h_{1,0}\\
h_{2,3} & h_{2,2} & h_{2,1} & h_{2,0} & 0 & 0\\
0 & h_{2,3} & h_{2,2} & h_{2,1} & h_{2,0} & 0\\
0 & 0 &  h_{2,3} & h_{2,2} & h_{2,1} & h_{2,0}
\end{array}\right]$$
is nonzero.
For $j\geq 5$, $Res(h_1,h_2)$ is a factor of a nontrivial fullsize minor of $\mathfrak{H}_j$. Hence, that all nontrivial fullsize minors of $\mathfrak{H}_j$ are nonzero implies that the resultant is nonzero. Therefore, it is enough to consider the nontrivial fullsize minors of $\mathfrak{H}_j$ to show complete $j$-MDP without having to take care of the resultant. However, for $j\leq 4$, one has to make sure that also the resultant is nonzero.

The probability of finding this type of codes increases for larger fields, so after randomly looking for $(2,1,3)$ complete $4$-MDP convolutional codes over $\mathbb{F}_{64}$ without success we were able to find $4$ such codes over $\mathbb{F}_{128}$. 

\begin{example} Let $H$ be the partial parity-check matrix of a $(2,1,3)$ complete $4$-MDP convolutional code over a finite field of characteristic $2$. Then 
\begin{itemize}
	\item the number of all full size minors of $H$ is $4368$;
	\item the number of different minors of $H$ is $1739$;
	\item the number of irreducible factors is $571$.
\end{itemize}
If the finite field is $\mathbb{F}_{128}$ then any of the following examples gives us $(2,1,3)$ complete $4$-MDP convolutional codes, since the corresponding resultants are nonzero.
\begin{itemize}
	\item $H_3=[1\quad 1]$, $H_2=[\alpha^6+\alpha^3\quad  \alpha^6+\alpha^5+\alpha^4+\alpha^2+1]$,  $H_1=[\alpha^5+\alpha^4\quad  1]$ and $H_0=[\alpha^4+\alpha+1\quad \alpha^4+\alpha^3+\alpha^2+1]$, with resultant $\alpha^2+\alpha+1\neq 0$;
	\item  $H_3=[1\quad 1]$, $H_2=[\alpha^4+\alpha^3\quad  \alpha^5]$, $H_1=[\alpha^6+\alpha^5+\alpha^2+\alpha\quad  \alpha^4+\alpha^3+\alpha^2]$ and $H_0=[\alpha^6+\alpha^4+\alpha^2+\alpha+1\quad  \alpha^3+\alpha^2+\alpha+1]$, with resultant $\alpha^6+\alpha^5+\alpha^4+\alpha^3+\alpha\neq 0$; 
	\item $H_3=[1\quad 1]$, $H_2=[\alpha^5+\alpha^3+\alpha^2+1\quad  \alpha^6+\alpha^5+\alpha^4+\alpha+1]$, $H_1=[\alpha^5+\alpha^4+\alpha^3+\alpha^2+\alpha\quad \alpha^4+\alpha^3+\alpha+1]$ and $H_0=[\alpha^6+\alpha^4+\alpha^3+\alpha^2+1\quad \alpha^3+\alpha]$, with resultant $\alpha^5+\alpha^4\neq 0$;
	\item $H_3=[1\quad 1]$, $H_2=[\alpha^6+\alpha^5+\alpha^4+\alpha+1\quad \alpha^6+\alpha^5+\alpha^3+\alpha^2+\alpha]$, $H_1=[\alpha^6+\alpha^5+\alpha^4\quad \alpha^5+\alpha^4+\alpha^3+\alpha^2+1]$ and $H_0=[\alpha^3+\alpha+1\quad \alpha^4+\alpha^2+1]$, with resultant $\alpha^5+\alpha^4+\alpha^3+1\neq 0$. 
\end{itemize}
\end{example}  

For a $(2,1,3)$ complete $5$-MDP convolutional code the number of irreducible factors is $1526$, but most of these factors have large expressions. So a randomized computer search, using the same algorithm, would take a few weeks to obtain a solution. In a $(2,1,3)$ complete MDP convolutional code we have $17731$ irreducible factors and we were able to store them in $89$ sets, but we need new ideas in order to deal with this amount of information faster.

\section{Description of Maple Algorithms and Future Work}\label{Maple}

In this section we give a description of the Maple algorithms and present a table with all results about the minimum field size of a $(2,1,2)$ complete $j$-MDP convolutional codes, for $j=1, 2, 3, 4$ (see fig. \ref{tab_all}). 

Our Maple algorithms for $(2,1,2)$ complete $j$-MDP convolutional codes has the following steps: For each $k\in\{2,3,4,5\}$,

\begin{enumerate}
	\item calculate all nonzero full size minors, and find out how many are different, of the $k\times 14$ submatrices of \eqref{Matrix5_14}, where $H_2=[1,1]$, $H_1=[a,b]$, $H_0=[c,d]$, and $a, b, c, d$ are free variables.  We may consider $H_2=[1,1]$ because the nonzeroness of a minor does not change if we multiply a column by a nonzero value.
	\item find all irreducible factors of all full size minors of the $k\times 14$ submatrices of \eqref{Matrix5_14}. Since we want to prove that all full size minors are nonzero, it is enough to prove that all the irreducible factors are nonzero.
	\item consider a finite field $\mathbb{F}$ and test if the irreducible factors are nonzero for any $a, b, c, d\in\mathbb{F}\setminus\{0\}$. For example, for $\mathbb{F}=\mathbb{F}_{16}$ we tested $15^4$ possible combinations of $a, b, c$ and $d$.
	\item to obtain complete descriptions we looked to all values of $a, b, c$ and $d$ that make all irreducible factors nonzero and figured out a formula to describe them. Then used Maple to show that the formula gives exactly all the values $a, b, c$ and $d$ obtained.
\end{enumerate}

A summary of the results obtained is illustrated in Figure \ref{tab_all}. For each $j\in\{1,2,3,4\}$ and for the smallest binary and prime field, we present:
\begin{itemize}
	\item the number of all full size minors,
	\item the number of different minors,
	\item the number of irreducible factors,
	\item the number of solutions and an example or the complete description.
	\item the percentage of solutions in relation to all the possible (nonzero) values for $a, b, c$ and $d$.
\end{itemize}

\begin{figure}[h!]\caption{Minimum field sizes for $(2,1,2)$ complete $j$-MDP convolutional codes, with $j=1, 2, 3, 4$.}
\begin{center}
\vspace{7mm}
\begin{tabular}{c|c|c|c|c|c|c|}\label{tab_all}
$j$ & Field & All  & Diff. & Irred. & Solutions & Perc.\\ \hline
$1$ & $\mathbb{F}_5$ & $91$ & $19$ & $10$ & $20$ & $7.812\%$\\
& & & & & $1,2,1,2$ & \\ \hline
& $\mathbb{F}_8$ & $91$ & $19$ & $10$ & $714$ & $29.738\%$\\
& & & & & $1,\alpha, \alpha+1, 1$ & \\ \hline \hline
$2$ & $\mathbb{F}_7$ & $364$ & $59$ & $21$ & $14$ & $1.080\%$\\
& & & & & $1,2,5,5$ & \\ \hline
& $\mathbb{F}_8$ & $364$ & $59$ & $19$ & $126$ & $5.248\%$\\
& & & & &  $1,\alpha, \alpha+1, \alpha+1$ & \\ \hline \hline
$3$ & $\mathbb{F}_{13}$ & $1001$ & $179$ & $44$ & $240$ & $1.157\%$\\
& & & & & $1,2,6,6$ & \\ \hline
& $\mathbb{F}_{16}$ & $1001$ & $179$ & $42$ & $600$ & $1.185\%$\\
& & & & & $1,\alpha, \alpha+1, \alpha+1$ & \\ \hline \hline
$4$ & $\mathbb{F}_{13}$ & $2002$ & $519$ & $83$ & $24$ & $0.115\%$\\
& & & & & $1,12,2,2$ & \\
& & & & & $2^{i_1},2^{i_2}, 2^{i_3}, 2^{i_4}$  & \\
& & & & & $i_1=0,\dots, 11$ & \\
& & & & & $i_2=i_1+6$  & \\
& & & & & $i_3=2i_1+1+6\ell$  & $\ell=0,1$ \\
& & & & & $i_4=i_3$  & \\ \hline
& $\mathbb{F}_{16}$ & $2002$ & $519$ & $79$ & $120$ & $0.237\%$\\
& & & & & $\alpha^{i_1},\alpha^{i_2}, \alpha^{i_3}, \alpha^{i_4}$  & \\
& & & & & $i_1=0,\dots, 14$ & \\
& & & & & $i_2=i_1+3k$  & $k=1,\dots, 4$\\
& & & & & $i_3=i_1+i_2-4^jk$  & $j=1,2$\\
& & & & & $i_4=i_3$  & \\ \hline \hline
\end{tabular}
\end{center}
\end{figure}

In the Maple algorithms for $(2,1,3)$ complete $4$-MDP convolutional codes we also found  all irreducible factors of all full size minors of the $5\times 16$ submatrix formed by the first $5$ rows and first $16$ columns of the matrix \eqref{Matrix7_20}. Then we did a randomized computer search of elements $a, b, c,d, e, f\in\mathbb{F}_{128}$ such that all these irreducible factors are nonzero. For the four examples exhibited in section \ref{comp213} we then calculated the corresponding resultant and fortunately found that it was nonzero for all cases.

In our future work we intent to study also complete $j$-MDP convolutional codes with other rates. A next step would be to consider rate $1/3$ or $2/3$. In particular, if the maximal delay for decoding is $2$, $(3,1,2)$ complete $2$-MDP convolutional codes respective $(3,2,2)$ complete $2$-MDP convolutional codes are the optimal choice and it seems reasonable to find such codes with the methods we used for $(2,1,2)$ complete MDP convolutional codes.
%


\section*{Acknowledgement}

Both authors are partially supported by the Portuguese Foundation for Science and Technology (FCT-Funda\c{c}\~{a}o para a Ci\^{e}ncia e a Tecnologia), through CIDMA - Center for Research and Development in Mathematics and Applications, by the projects with references UIDB/04106/2020 and UIDP/04106/2020. Moreover, the second listed author gratefully acknowledges the support from the German Research Foundation within grant LI3101/1-1.

\bibliography{mybibfile}

\begin{thebibliography}{}

\bibitem{gj}
G. N. Alfarano and J. Lieb, A simplified criterion for MDP convolutional codes, arXiv:2003.07322.



\bibitem{dr16}
P. J. Almeida, D. Napp and R. Pinto, Superregular matrices and applications to convolutional codes, {\it Linear Algebra Appl.} \textbf{499} (2016) 1--25.



\bibitem{dr13}
P. J. Almeida, D. Napp and R. Pinto, A new class of superregular matrices and MDP convolutional codes, {\it Linear Algebra Appl.} \textbf{439} (2013) 2145--2157.



\bibitem{bar}
A. Barbero and O. Ytrehus, Rate $(n-1)/n$ Systematic Memory Maximum Distance Separable Convolutional Codes, {\it IEEE Transactions on Information Theory} \textbf{64(4)} (2018) 3018--3030.


\bibitem{strongly}
H. Gluesing-Luerssen, J. Rosenthal and R. Smarandache, Strongly-MDS Convolutional Codes, {\it IEEE Transactions on Information Theory} \textbf{52.2} (2006) 584--598.




\bibitem{h}
R. Hutchinson, The existence of strongly MDS convolutional codes, {\it SIAM J. Control Optim.} \textbf{47} (2008) 2812--2826.


\bibitem{mdp} R. Hutchinson, J. Rosenthal and R. Smarandache, Convolutional codes with maximum distance profile, {\it Systems $\&$ Control Letters} \textbf{54} (2005) 53--63.






\bibitem{b} R. Hutchinson, R. Smarandache and J. Trumpf, On superregular matrices and MDP convolutional codes, {\it Linear Algebra Appl.}  \textbf{428} (2008) 2585--2596.

\bibitem{cmdp} J. Lieb, Complete MDP convolutional codes, {\it J. Algebra App.} \textbf{18} no. 6 (2019), page 1950105.



\bibitem{nf} J. Lieb, Necessary Field Size and Probability for MDP and Complete MDP Convolutional Codes, {\it Designs, Codes and Cryptography} \textbf{87} (2019) 3019--3043.




\bibitem{dr}
D. Napp and R. Smarandache, Constructing strongly MDS convolutional codes with maximum distance profile, {\it Advances in Mathematics of Communications} \textbf{10(2)} (2016) 275--290.






\bibitem{RS99}
J. Rosenthal and R. Smarandache, Maximum distance separable convolutional codes, {\it Appl. Algebra Engrg. Comm. Comput.} \textbf{10} (1999) 15--32.

\bibitem{ro01} J. Rosenthal. Connections between linear systems and convolutional codes. In: B.  Marcus  and  J.  Rosenthal,  editors, {\it Codes, Systems and Graphical Models}, IMA Vol. 123, pages 39–66. Springer-Verlag, 2001.


\bibitem{vp} V. Tomas, J. Rosenthal and R. Smarandache, Decoding of Convolutional Codes Over the Erasure Channel, {\it IEEE Transactions on Information Theory} \textbf{58(1)} (2012) 90--108.


\end{thebibliography}

\appendix

\section{Maple Programs}

The next instructions produce all the irreducible factors of all fullsize minors of the parity-check matrix of a $(2,1,2)$ complete MDP convolutional code over a binary field. Afterwards all solutions over $\mathbb{F}_{16}$ are produced. A similar code can be obtained over other prime fields.

\begin{verbatim}

restart:
with(LinearAlgebra):with(combinat):
A:=Matrix([[1,1,a,b,c,d,0,0,0,0,0,0,0,0],[0,0,1,1,a,b,c,d,0,0,0,0,0,0],
[0,0,0,0,1,1,a,b,c,d,0,0,0,0],[0,0,0,0,0,0,1,1,a,b,c,d,0,0],
[0,0,0,0,0,0,0,0,1,1,a,b,c,d]]):
setminors:={}:
U:=choose(14,5):
for k to nops(U) do
B:=SubMatrix(A,[seq(i,i=1..5)],U[k]):
detB:=Determinant(B):
setminors:=setminors union {detB} mod 2:
od:
setminorsaux:=setminors minus {0,1}:
irredfact:={}:
for iminors in setminorsaux do
irred:=Factors(iminors) mod 2:
for irri to nops(irred[2]) do
irredfact:=irredfact union {irred[2][irri][1]} mod 2:
od:
od:
print(irredfact):print(nops(setminors),nops(irredfact)):


functset:=(a,b,c,d)->{a, b, c, d, a+b, a^2+c, b^2+d, a*b+c, 
	a*b+d, a*d+b*c, a^4+a^2*c+c^2, b^4+b^2*d+d^2, a^2*b+a*c+b*c, 
	a^3*b+a^2*c+c^2, a*b^2+a*d+b*d, a*b^3+b^2*d+d^2, a*b+c+d, 
	a*b^2+a*d+b*c, a^2*b+a*d+b*c, a*b^3+b^2*c+c*d, a^2*b^2+a*b*d+c^2, 
	a^2*b^2+a*b*c+d^2, a^3*b+a^2*d+c*d, a^4*b+a^3*c+a^2*b*c+b*c^2, 
	a*b^4+a*b^2*d+b^3*d+a*d^2, a*b+b^2+c+d, a^2+a*b+c+d, 
	a*b*d+b^2*c+c*d+d^2, a*b^2+a*d+b*c+b*d, a*b^2+b^3+a*d+b*c, 
	a^2*d+a*b*c+c^2+c*d, a^2*b+a*c+a*d+b*c, a^2*b+a*b^2+a*c+b*d, 
	a^3+a^2*b+a*d+b*c, a*b^2*d+b^3*c+a*d^2+b*d^2, a*b^3+b^2*c+b^2*d+d^2, 
	a*b^3+b^2*c+b^2*d+c*d, a^2*b*d+a*b^2*c+a*d^2+b*c^2, 
	a^2*b^2+c^2+c*d+d^2, a^3*d+a^2*b*c+a*c^2+b*c^2, a^3*b+a^2*c+a^2*d+c*d, 
	a^3*b+a^2*c+a^2*d+c^2, a*b^4+a*b^2*d+b^3*c+a*d^2, 
	a^2*b^3+a*b^2*d+b^3*c+a*d^2, a^2*b^3+a*b^2*c+b^3*c+a*c*d, 
	a^3*b^2+a^3*d+a^2*b*d+b*c*d, a^3*b^2+a^3*d+a^2*b*c+b*c^2, 
	a^4*b+a^3*d+a^2*b*c+b*c^2, a*b^3+a*b*d+b^2*c+b^2*d+d^2, 
	a*b^3+a*b*d+b^2*c+b^2*d+c*d, a^2*b^2+a*b*c+a*b*d+b^2*c+c*d, 
	a^2*b^2+a^2*d+a*b*d+b^2*c+c*d, a^2*b^2+a^2*d+a*b*c+b^2*c+c*d, 
	a^2*b^2+a^2*d+a*b*c+a*b*d+c*d, a^3*b+a^2*c+a^2*d+a*b*c+c*d, 
	a^3*b+a^2*c+a^2*d+a*b*c+c^2, a^2*d+a*b*c+a*b*d+b^2*c+c^2+d^2, 
	a*b^3+b^4+b^2*c+b^2*d+c*d+d^2, a^2*b^2+a*b^3+b^2*c+b^2*d+c^2+c*d, 
	a^3*b+a^2*b^2+a^2*c+a^2*d+c*d+d^2, a^4+a^3*b+a^2*c+a^2*d+c^2+c*d, 
	a*b^3*d+b^4*c+b^2*c*d+b^2*d^2+c*d^2+d^3, 
	a^2*b^2*d+a*b^3*c+b^2*c^2+b^2*c*d+c*d^2+d^3, 
	a^2*b^3+a*b^2*c+a*b^2*d+a*c*d+a*d^2+b*c*d, 
	a^2*b^3+a^2*b*d+a*b^2*d+a*c*d+a*d^2+b*c^2, 
	a^2*b^3+a^2*b*d+a*b^2*c+a*c*d+a*d^2+b*d^2, 
	a^2*b^3+a^2*b*d+a*b^2*c+a*b^2*d+b^3*c+a*d^2, 
	a^2*b^3+a^2*b*d+a*b^2*c+a*b^2*d+b^3*c+a*c*d, 
	a^3*b*d+a^2*b^2*c+a^2*c*d+a^2*d^2+c^3+c^2*d, 
	a^3*b^2+a^2*b*d+a*b^2*c+a*c^2+b*c^2+b*c*d, 
	a^3*b^2+a^2*b*c+a*b^2*c+a*d^2+b*c^2+b*c*d, 
	a^3*b^2+a^2*b*c+a^2*b*d+a*c*d+b*c^2+b*c*d, 
	a^3*b^2+a^3*d+a^2*b*c+a^2*b*d+a*b^2*c+b*c*d, 
	a^3*b^2+a^3*d+a^2*b*c+a^2*b*d+a*b^2*c+b*c^2, 
	a^4*d+a^3*b*c+a^2*c^2+a^2*c*d+c^3+c^2*d, 
	a^2*b^2+a*b^3+a^2*d+a*b*c+a*b*d+b^2*d+c*d+d^2, 
	a^3*b+a^2*b^2+a^2*c+a*b*c+a*b*d+b^2*c+c^2+c*d, 
	a^2*b^2*d+a*b^3*c+a^2*d^2+a*b*c*d+a*b*d^2+b^2*c^2+c^2*d+c*d^2, 
	a^3*b*d+a^2*b^2*c+a^2*d^2+a*b*c^2+a*b*c*d+b^2*c^2+c^2*d+c*d^2}:
irredfactval:=functset(a1,b1,c1,d1):
alias(alpha = RootOf(x^4+x+1=0,2)):
interv:=[17,31]:
for elementa from interv[1] to interv[2] do
base2a:=convert(elementa,base,2): 
a1:=base2a[1]*1+base2a[2]*alpha+base2a[3]*alpha^2+base2a[4]*alpha^3:
for elementb from interv[1] to interv[2] do
base2b:=convert(elementb,base,2): 
b1:=base2b[1]*1+base2b[2]*alpha+base2b[3]*alpha^2+base2b[4]*alpha^3:
for elementc from interv[1] to interv[2] do
base2c:=convert(elementc,base,2): 
c1:=base2c[1]*1+base2c[2]*alpha+base2c[3]*alpha^2+base2c[4]*alpha^3:
for elementd from interv[1] to interv[2] do
base2d:=convert(elementd,base,2): 
d1:=base2d[1]*1+base2d[2]*alpha+base2d[3]*alpha^2+base2d[4]*alpha^3:
pass:=true:
irredfactf:={}:
for fink in irredfactval do
irredfactf:=irredfactf union {evala(fink) mod 2}:
if evala(fink) mod 2=0 then pass:=false:#print(fink):
fi:
od:
if pass then print(a1,b1,c1,d1):
print(nops(irredfactf)):
setsol:=setsol minus {[a1,b1,c1,d1]}:
fi:
od:
od:
od:
od:
\end{verbatim}

For the randomized computer search of $(2,1,3)$ complete $4$-MDP convolutional codes over $\mathbb{F}_{128}$ we use the following algorithm, after creating $6$ sets {\bf irredfactval} where we put all the $571$ irreducible factors.

\begin{verbatim}
p:=2:q:=p^7:
nv:=q^6-1:
with(RandomTools[MersenneTwister]):
count:=0:
alias(alpha = RootOf(x^7+x+1=0,2)):
for tries to 100000 do
element:=GenerateInteger() mod nv:
basen:=convert(element,base,q):
num:=nops(basen):
if num>=6 then
basea:=convert(basen[1]+q+1,base,2):
a1:=basea[1]+basea[2]*alpha+basea[3]*alpha^2+basea[4]*alpha^3+basea[5]*alpha^4+
basea[6]*alpha^5+basea[7]*alpha^6:
baseb:=convert(basen[2]+q+1,base,2):
b1:=baseb[1]+baseb[2]*alpha+baseb[3]*alpha^2+baseb[4]*alpha^3+baseb[5]*alpha^4+
baseb[6]*alpha^5+baseb[7]*alpha^6:
basec:=convert(basen[3]+q+1,base,2):
c1:=basec[1]+basec[2]*alpha+basec[3]*alpha^2+basec[4]*alpha^3+basec[5]*alpha^4+
basec[6]*alpha^5+basec[7]*alpha^6:
based:=convert(basen[4]+q+1,base,2):
d1:=based[1]+based[2]*alpha+based[3]*alpha^2+based[4]*alpha^3+based[5]*alpha^4+
based[6]*alpha^5+based[7]*alpha^6:
basee:=convert(basen[5]+q+1,base,2):
e1:=basee[1]+basee[2]*alpha+basee[3]*alpha^2+basee[4]*alpha^3+basee[5]*alpha^4+
basee[6]*alpha^5+basee[7]*alpha^6:
basef:=convert(basen[6]+q+1,base,2):
f1:=basef[1]+basef[2]*alpha+basef[3]*alpha^2+basef[4]*alpha^3+basef[5]*alpha^4+
basef[6]*alpha^5+basef[7]*alpha^6:
pass:=true:
irredfactf:={}:
for setk to 6 do
for fink in irredfactval[setk] do
irredfactf:=irredfactf union {evala(fink) mod p}:
if evala(fink) mod p=0 then pass:=false:#print(fink):
fi:
od:
if pass and setk>1 then print(a1,b1,c1,d1,e1,f1,setk):
fi:
od:
if pass then print(a1,b1,c1,d1,e1,f1):count:=count+1:
print(nops(irredfactf)):
fi:
fi:
od:
\end{verbatim}

\end{document}